\documentclass[aip,jcp,twocolumn]{revtex4}

\usepackage{amssymb}
\usepackage{amsmath}
\usepackage{amsfonts}
\usepackage{graphicx}
\usepackage{bbm}

\usepackage{hyperref} 

\hypersetup{
    colorlinks=true,
    linkcolor=blue,
    citecolor=blue,
    filecolor=blue,
    urlcolor=blue,
}

\newcommand{\unit}[1]{\mathbf{\hat{#1}}}
\newcommand{\de}[1]{\delta #1}

\begin{document}

\title{Models for twistable elastic polymers in Brownian dynamics, 
and their implementation for \textsc{LAMMPS}}

\author{C. A. Brackley}
\author{A. N. Morozov}
\author{D. Marenduzzo}
\affiliation{SUPA, School of Physics and Astronomy, University of Edinburgh, Mayfield Road, Edinburgh, EH9 3JZ, UK}

\begin{abstract}
An elastic rod model for semi-flexible polymers is presented. Theory for a continuum rod is reviewed, and it is shown that a popular discretised model used in numerical simulations gives the correct continuum limit. Correlation functions relating to both bending and twisting of the rod are derived for both continuous and discrete cases, and results are compared with numerical simulations. Finally, two possible implementations of the discretised model in the multi-purpose molecular dynamics software package \textsc{lammps} are described.
\end{abstract}

\maketitle 

\section{Introduction}

When setting up a coarse grained model for a polymer or biopolymer
it is often useful, if not necessary, to account for its ability to both
bend and twist. A notable example is that of double-stranded DNA.
A first coarse grained description of DNA is to view it
as a semi-flexible polymer, or a worm-like chain~\cite{wlc1}, 
i.e. a fluctuating elastic rod with a bending rigidity, 
and an associated persistence length. The latter gives a measure of the 
length scale along the polymer backbone over which 
correlations in the local direction (the tangent to the backbone) decay. 
For DNA in a physiologically realistic salt solution this length is 50 nm, which is significantly larger than its  thickness (2.5 nm for hydrated B-DNA). 

While a worm-like chain provides a simple and useful 
description of DNA under some conditions~\cite{wlc1,wlc2,wlc3,wlc4,wlc5},
it neglects a key elastic property of this biomolecule: its resistance to
twisting.
Indeed, in reality DNA is a double helix which in its relaxed state has a pitch of around 10 base pairs (in
B-DNA), and over-twisting or under-twisting the helix incurs an energetic
penalty~\cite{understandingdna,marko1}. 
Twisting DNA may also lead to supercoiling~\cite{supercoilingreview,marko2,Forth2008}, 
i.e. the writhing of a highly twisted filament, which can be witnessed in 
the everyday world when twisting up an office telephone cord.
Supercoiling can be due to over or under-twisting the helical DNA,
and is accordingly referred to as either positive or negative.

Supercoiling is important for DNA organisation and function within cells. 
Within bacteria, supercoiling helps pack the genome into the 
tight volume of the cell~\cite{bacterialsupercoiling1}, 
and Hi-C experiments suggest that this phenomenon is also gives rise to the
contact maps observed for the bacterial chromosome~\cite{mirnybacteria}.
Furthermore, it was realised many years ago that supercoiling is
naturally and generically created {\it in vivo}~\cite{twinsupercoiled,transcription,Deng2005}: as an RNA polymerase transcribes a gene, it generates an excess of twist ahead, 
and a deficit behind, and this leads to, respectively, positively and negatively 
supercoiled domains (a Brownian dynamics simulation exploring the biophysics of 
this phenomenon was provided in~\cite{twinsupercoiledBD}).
Supercoiling is also thought to play a functional role in gene
regulation~\cite{supercoilingnick} and transcription~\cite{supercoilingchromatin} 
in eukaryotes.
This is because, among other things, as the DNA wraps around histones 
to make chromatin, each wrapping adds two units of writhe into the polymer: 
the interplay of this writhe with the transcriptionally-driven supercoiling 
just discussed could help open up the chromatin fibre as it is transcribed 
(or replicated)~\cite{supercoilingchromatin,supercoilingreview,Roca2011,Mondal2001}.
The importance of supercoiling to intracellular DNA is also apparent
from the number of key enzymes whose role is to regulate it, such
as DNA gyrases, which induce negative supercoiling, and topoisomerases,
which are often employed to relieve it~\cite{supercoilingreview}.
 
It is therefore vital for coarse grained simulations of 
bacterial and eukaryotic DNA, or of naked DNA loops, to be able
to treat twist and twist fluctuations accurately. The simplest way to
achieve this is to endow the model DNA with a twist rigidity as well
as a bending rigidity~\cite{marko1,Bouchiat2000}. Just as the bending rigidity leads 
to a non-zero persistence length which characterises how correlations in
the backbone directions decay due to bending, the twist rigidity leads to
a twist persistence length, which characterises how twist correlations
decay, or, equivalently, how big twist fluctuations should be at a given 
temperature $T$. This twist persistence length is of the same order
of, and slightly larger than, the bending persistence length, 
and typical estimates are between 60 and 75 nm~\cite{marko2,Fujimoto1990,Lipfert2010}.

There are many excellent papers in the literature which discuss how
to set up a model of a twistable worm-like chain to study
the dynamics of DNA when supercoiling is 
important~\cite{powersreview,twistsim1,twistsim2,twistsim3,depablo,schlick1,schlick2,schlick3,rappaport2}. 
Most of the Brownian dynamics simulation work on coarse grained (bead-and-spring) DNA molecules with twisting 
and supercoiling build on the seminal contributions in 
Refs.~\cite{Chirico1994,Chirico1996,Allison1989}. However, most modern,
widely-used molecular dynamics codes, such as \textsc{lammps}~\cite{lammps}, do not 
currently incorporate this force field in their source code.
Furthermore, in the literature there have been a variety
of approaches to study twistable elastic chains, and to our knowledge
there has not been  a systematic analysis of the possible Hamiltonians,
and the relation between them and with the available continuum
theories. Thus our goal in this work is to fill this gap and provide
a detailed description of possible coarse grained Hamiltonians for
twistable worm-like chains, suitable for use in Brownian or Langevin dynamics simulations.
We also derive formulas for the tangent-tangent and for the
twist correlation functions, and describe several possible implementations 
of these models in \textsc{lammps}.

Our work is structured as follows. In the next Section, we will review the
continuum theory of elastic rods. In Sec.~III, we discretise the continuum energy and show that the model in 
Ref.~\cite{Chirico1994} provides a valid discretisation, as do other
equally valid options which are formulated using appropriate combinations of dihedral 
potentials. Section IV
contains an analytical calculation of the persistence lengths in the discrete
model, together with a comparison with numerical data. Section V
provides a detailed description of the implementation of some of the
models introduced in Sec.~III in \textsc{lammps}, and then in Sec.~VI we compare the some simulations results with the theory. Finally, in Sec.~VII we draw some conclusions, and point to some possible ways in which the current
work may be applied to DNA biophysics.

\section{A Continuous Elastic Rod}\label{continuous}

\begin{figure}
\includegraphics{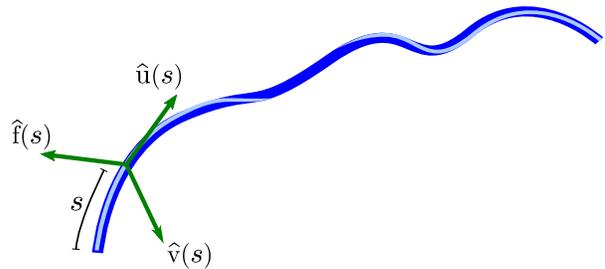}
\caption{\label{fig:rod} Schematic representation of an elastic rod showing the Darboux frame at a point $s$ along the rod.}
\end{figure}

A non-extensible elastic rod can be described mathematically as a ``stripe'' (a thin slice of a plane), whose position in space is given by the vector $\mathbf{r}(s)$ where $s$ is the distance along the rod~\cite{Rappaport2007}. 
A continuous curve in three dimensional space can be described by the Frenet-Serret frame, made up of the vector tangent to the curve $\unit{u}(s)=d\mathbf{r}(s)/ds$, a normal vector $\unit{n}$ which points in direction of  $d\unit{u}/ds$, and a binormal $\unit{b}=\unit{u}\times\unit{n}$; the relationship between these vectors and their derivatives (i.e. a description of the rotation of the frame as one moves along the curve) is given by the Frenet-Serret equation (see Ref.~\cite{Rappaport2007}). Whilst a curve is uniquely defined by the tangent (the other two vectors are constructed from this), a stripe with a finite thickness requires two vectors to define it: the tangent $\unit{u}(s)$ and the local normal to the plane $\unit{f}(s)$ (also known as the material normal). The configuration of a stripe is described by the Darboux frame, consisting of $\unit{u}(s)$, $\unit{f}(s)$ and a vector $\unit{v}(s)$ perpendicular to these 
(defined such that $\unit{f}\times\unit{v}=\unit{u}$, so that $\unit{f}$, $\unit{v}$ and $\unit{u}$  correspond respectively to the usual $x$, $y$ and $z$ axes); a schematic representation is given in Fig.~\ref{fig:rod}.  
The Frenet-Serret and Darboux frames are related by
\[
\left( \begin{matrix}
\unit{u} \\ \unit{f} \\ \unit{v}
\end{matrix} \right) = S \left( \begin{matrix}
\unit{u} \\ \unit{n} \\ \unit{b}
\end{matrix}\right)
\]
where 
\[
S=\left(\begin{matrix}
1 & 0 & 0 \\
0 & \cos\theta & \sin\theta \\
0 & -\sin\theta & \cos\theta
\end{matrix}\right)
\]
with $\theta$ the rotation angle between the two frames.

The path of the rod is represented by the generalized curvatures $\omega_i(s)$ which determine the unit vectors by the \textit{generalized} Frenet-Serret~\cite{Rappaport2007} equation
\begin{equation}\label{GFSeq}
\frac{d}{ds} \left( \begin{matrix} \unit{u} \\ \unit{f} \\ \unit{v} \end{matrix} \right) = 
\left( \begin{matrix} 
0 & \omega_2 & -\omega_1 \\
-\omega_2 & 0 & \omega_3 \\
\omega_1 & -\omega_3 & 0 \\
\end{matrix} \right)
\left( \begin{matrix} 
\unit{u} \\ \unit{f} \\ \unit{v}
\end{matrix} \right),
\end{equation}
where it can be seen that $\omega_3(s)ds$ is the infinitesimal angle of rotation about the direction $\unit{u}(s)$, $\omega_1(s)ds$ is the infinitesimal rotation about $\unit{f}(s)$ etc. The in-extensibility of the rod can be expressed by the condition $|d\mathbf{r}/ds|=1$.
The energy for an elastic rod is given by
\[
E_{\rm ER} = \frac{1}{2} \int_0^L \left( \kappa_{b1}\omega_1(s)^2 + \kappa_{b2}\omega_2(s)^2 + \kappa_t\omega_3(s)^2 \right) ds,
\]
where $\kappa_{b1}$ and $\kappa_{b2}$ are bending rigidities and $\kappa_t$ is the twist rigidity (each having units of energy$\cdot$length). If the elastic rod has a circular cross section, then $\kappa_{b1}=\kappa_{b2}=\kappa_b$, giving
\begin{equation}
E_{\rm ER} = \frac{1}{2} \int_0^L \left( \kappa_b(\omega_1^2+\omega_2^2) + \kappa_t\omega_3^2 \right) ds.
\label{e_er_omegas}
\end{equation}

By inspection of Eq.~(\ref{GFSeq}) we can identify the curvature of the rod
\[
\omega_1(s)^2+\omega_2(s)^2 = \left| \frac{d\unit{u}(s)}{ds}\right|^2,
\]
and also
\[
\omega_3(s)^2 = \unit{u}(s) \cdot \frac{d\unit{f}(s)}{ds} \times \frac{d\unit{v}(s)}{ds}.
\]
One can therefore also write the energy
\begin{align}\label{continuum}
E_{\rm ER} = \frac{1}{2} &\int_0^L \Bigg[ \kappa_b\left| \frac{d\unit{u}(s)}{ds}\right|^2 \nonumber \\
&+  \kappa_t~\unit{u}(s) \cdot \left(\frac{d\unit{f}(s)}{ds} \times \frac{d\unit{v}(s)}{ds}\right) \Bigg] ds.
\end{align}

The flexibility of the rod can be described by looking at correlations in the direction of the vectors along the rod. For example, the bending correlations are given by 
$\langle \unit{u}(0) \cdot \unit{u}(s) \rangle$,
whereas the twist correlations are given by 
$\langle \unit{f}(0) \cdot \unit{f}(s) \rangle$. 
We include a simple calculation of these expressions,
in the continuous limit, in Appendix~\ref{app:corfun}.

\section{Discrete Representation of an Elastic Rod}\label{discrete}

We now consider a discrete elastic rod model, with segments of length $a$ which are defined by the position vectors of the vertices between each segment $\mathbf{r}_i$, $i=1,\dots N$. The configuration of the rod can be described by the tangents $\unit{u}_i=(\mathbf{r}_{i+1}-\mathbf{r}_i)/a$~\cite{foot_N}, and the orthogonal vectors $\unit{f}_i$ and $\unit{v}_i$. Together $\unit{u}_i,\unit{f}_i$ and $\unit{v}_i$ make up a frame of reference on vertex $i$.

To discretise the energy, we start from the functional in
Eq.~(\ref{continuum}). By using the following discretised approximations
\begin{eqnarray}
\frac{d\unit{u}(s)}{ds} & = & \frac{\unit{u}_{i+1}-\unit{u}_i}{a} \nonumber\\ 
\frac{d\unit{f}(s)}{ds} & = & \frac{\unit{f}_{i+1}-\unit{f}_i}{a} \nonumber\\ 
\frac{d\unit{v}(s)}{ds} & = & \frac{\unit{v}_{i+1}-\unit{v}_i}{a} \nonumber
\end{eqnarray}
we obtain
\begin{align}
E_{\rm DER}  = &  \frac{\kappa_b}{2} \sum_{i=1}^{N-1} \frac{(\unit{u}_{i+1}-\unit{u}_i)^2}{a}  \nonumber \\
&+\frac{\kappa_t}{2} \sum_{i=1}^{N-1} \unit{u}_i\cdot \frac{(\unit{f}_{i+1}-\unit{f}_i)\times (\unit{v}_{i+1}-\unit{v}_i)}{a} \nonumber \\ \label{discretev2}
  = &  \frac{\kappa_b}{a} \sum_{i=1}^{N-1} \left(1-\unit{u}_{i+1}\cdot\unit{u}_{i}\right) \nonumber \\
&+ \frac{\kappa_t}{2a} \sum_{i=1}^{N-1} 
\left(1+\unit{u}_{i+1}\cdot\unit{u}_{i}-\unit{f}_{i+1}\cdot\unit{f}_{i}
-\unit{v}_{i+1}\cdot\unit{v}_{i}\right).
\end{align}
To get the final formula, we have used the following mixed product
identity, 
\begin{eqnarray}
\mathbf{a}\cdot\left(\mathbf{b}\times\mathbf{c}\right) = 
\mathbf{b}\cdot\left(\mathbf{c}\times\mathbf{a}\right) =
\mathbf{c}\cdot\left(\mathbf{a}\times\mathbf{b}\right)
\end{eqnarray}
which holds for any triplet of vectors, $\mathbf{a}$, $\mathbf{b}$
and $\mathbf{c}$.

Another useful parametrisation of the energy is to use Euler angles, where a set of three angles $\alpha$, $\beta$, and $\gamma$ is used to describe a rotation from one frame of reference to another. Here we use the $z\mbox{-}x'\mbox{-}z''$ Euler angle convention where the first angle is a rotation of the initial frame about the $z$-axis, the second a rotation about the new $x$-axis, and the third about the new $z$-axis. Rotations have positive sign if they are clockwise when looking along the axis. The rotation can be written as a matrix $\mathcal{R}$, for example to rotate the reference frame on vertex $i$ to that on $i+1$, 
$\left( \unit{u}_{i+1},\unit{f}_{i+1},\unit{v}_{i+1}\right)^{\rm T} = \mathcal{R} \left( \unit{u}_{i},\unit{f}_{i},\unit{v}_{i}\right)^{\rm T}$
The matrix can be decomposed into individual rotations 
\begin{equation}\label{decomposedRot}
\mathcal{R} = R_{\unit{u}}(\gamma)R_{\unit{f}}(\beta)R_{\unit{u}}(\alpha),
\end{equation}
where
\begin{equation}
R_{\unit{u}}(\theta)=\left( \begin{matrix}
1 & 0 & 0 \\
0 & \cos\theta & \sin\theta \\
0 & -\sin\theta & \cos\theta 
\end{matrix}\right),\nonumber \\
\end{equation}
and
\begin{equation}
R_{\unit{f}}(\theta)=\left( \begin{matrix}
\cos\theta & 0 & -\sin\theta \\
0 & 1 & 0 \\
\sin\theta & 0 & \cos\theta
\end{matrix} \right),\nonumber
\end{equation}
are rotation matrices such that e.g. $R_{\unit{f}}(\theta) \left( \unit{u},\unit{f},\unit{v} \right)^{\rm T}$ gives a set of axis vectors which have been rotated by an angle $\theta$ about $\unit{f}$.

By using angles, $\alpha_i$, $\beta_i$ and $\gamma_i$ to describe the rotation between vertex $i$ and $i+1$,
Eq.~(\ref{discretev2}) can be rewritten as
\begin{align}\label{discreteeulerv1}
E_{\rm DER} &=  \frac{\kappa_b}{a} \sum_{i=1}^{N-1} \left[1-\cos{\beta_i}\right] \nonumber\\
&~+ \frac{\kappa_t}{a} \sum_{i=1}^{N-1}\frac{\left(1+\cos{\beta_i}\right)}{2}
\left[1-\cos(\alpha_i+\gamma_i)\right].
\end{align}
Expanding this to second order in $\beta_i$ and $\alpha_i+\gamma_i$ [which
are both small due to the Boltzmann weight associated to the energy function
in Eq.~(\ref{discreteeulerv1})], we obtain
\begin{align}\label{chiricolangowski}
E_{\rm DER} &=  \frac{\kappa_b}{a} \sum_{i=1}^{N-1} \frac{\beta_i^2}{2} 
+\frac{\kappa_t}{a} \sum_{i=1}^{N-1}\frac{(\alpha_i+\gamma_i)^2}{2},
\end{align}
which is the energy function proposed in Ref.~\cite{Chirico1994}
and used in most subsequent works. This Hamiltonian is therefore 
equivalent to Eq.~(\ref{continuum}) in the continuum limit,
where $a\to 0$ while $\kappa_{b,t}/a$ are kept constant. In the
same limit, another viable version, which we will use in the rest of this work, is
\begin{equation}\label{Ediscreteeq}
E_{\rm DER} =  \frac{\kappa_b}{a} \sum_{i=1}^{N-1} \left[1-\cos{\beta_i}\right] + \frac{\kappa_t}{a} \sum_{i=1}^{N-1}\left[1-\cos(\alpha_i+\gamma_i)\right],
\end{equation}
where the $\beta_i$ and $\alpha_i+\gamma_i$ angles are treated symmetrically.

We note that the energy for each segment is independent, and Eq.~(\ref{Ediscreteeq}) can be written as a sum $E_{\rm DER} = \sum_i E_i$. A typical simulation of such a discrete rod would represent each vertex as a bead, employing a spring potential between each (e.g. a harmonic or FENE springs) to account for extension of the polymer, and a steric interaction potential to prevent beads overlapping. The total energy for the system would therefore be a sum of these components $\mathcal{H}=E_{\rm DER} + E_{\rm spring} + E_{\rm steric}$.

\section{Correlation functions for the discrete elastic rod}\label{sec:corr}

In this section we find correlation functions for bending and twisting along the discrete rod, similar to those described in Sec.~\ref{continuous} (see Appendix~\ref{app:corfun}) for the continuous case. We will therefore need to calculate the probability of a given chain configuration
\begin{equation}
\mathcal{P}( \{ \mathbf{r}_i \} ) = \prod_i \frac{1}{Z_i} e^{-E_i/k_BT},
\end{equation}
where $E_i$ is the $i$th term of the sums in Eq.~(\ref{Ediscreteeq}) and
\begin{equation}
Z_i=\int e^{-E_i/k_BT} \mathcal{D}(\alpha_i,\beta_i,\gamma_i),
\end{equation}
with $\mathcal{D}(\alpha,\beta,\gamma)$ the volume element for integrating over orientations parametrised by the Euler angles. To find the correlation functions it will useful to write down a rotation or ``transfer'' matrix, which describes the rotation of the frame at vertex $i$ required to get that at $i+1$~\cite{transfermatrix}. Using the Euler angle formulation this is simply the matrix given in Eq.~(\ref{decomposedRot}), i.e.
\begin{equation} \label{transfermatrixequation}
\left( \begin{matrix} \unit{u}_{i+1} \\ \unit{f}_{i+1} \\ \unit{v}_{i+1} \end{matrix} \right) =
\mathcal{R}
\left( \begin{matrix} \unit{u}_i \\ \unit{f}_i \\ \unit{v}_i \end{matrix} \right) 
\end{equation}
where
\begin{widetext}
\begin{equation}
\mathcal{R} = \left( \begin{matrix} 
\cos\beta_i & \sin\alpha_i \sin\beta_i & -\cos\alpha_i \sin\beta_i\\
\sin\beta_i \sin\gamma_i & \cos\gamma_i \cos\alpha_i - \cos\beta_i \sin\alpha_i \sin \gamma_i &  \sin\alpha_i \cos\gamma_i + \cos\beta_i\sin\gamma_i\cos\alpha_i\\
\sin\beta_i \cos\gamma_i & -\cos\alpha_i \sin\gamma_i - \cos\beta_i \cos\gamma_i \sin\alpha_i &-\sin\gamma_i \sin\alpha_i + \cos\beta_i\cos\alpha_i\cos\gamma_i 
\end{matrix} \right). 
\end{equation}
\end{widetext}
Alternatively we can write this as a set of equations for the $i+1$th set of vectors as functions of the $i$th
\begin{align}
\unit{u}_{i+1} =& \cos\beta_i \unit{u}_i + \sin\alpha_i\sin\beta_i \unit{f}_i - \cos\alpha_i\sin\beta_i \unit{v}_i, \nonumber\\ 
\unit{f}_{i+1} =& \sin\beta_i\sin\gamma_i \unit{u}_i + (\cos\gamma_i\cos\alpha_i - \cos\beta_i\sin\alpha_i\sin\gamma_i) \unit{f}_i  \nonumber\\
& + (\sin\alpha_i\cos\gamma_i + \cos\beta_i\sin\gamma_i\cos\alpha_i)  \unit{v}_i,\nonumber\\
 \unit{v}_{i+1} =&  \sin\beta_i\cos\gamma_i \unit{u}_i - (\cos\alpha_i\sin\gamma_i + \cos\beta_i\cos\gamma_i\sin\alpha_i) \unit{f}_i \nonumber\\
&- (\sin\gamma_i\sin\alpha_i - \cos\beta_i\cos\alpha_i\cos\gamma_i)  \unit{v}_i.
\label{vecip1i}
\end{align}

Correlation functions can be found from the Eigenvectors of the average of the transfer matrix $\langle \mathcal{R} \rangle$. The matrix elements are found, e.g. by integrating $\langle \cos\beta_i \rangle = \int \cos\beta_i \mathcal{P}( \{ \mathbf{r}_i \} ) \mathcal{D}(\alpha_i,\beta_i,\gamma_i)$. Since each segment decouples, i.e. $P(\{\mathbf{r}_i\})=\prod_i P_i(\mathbf{r}_i)$, we can consider averages on only an individual segment, and here-on drop the $i$ index where appropriate. To perform the integrals we note that $\beta$ is in the interval $[0,\pi]$, whereas $\alpha,\gamma\in[-\pi,\pi]$; the appropriate volume element is 
\begin{equation}
\mathcal{D}(\alpha,\beta,\gamma) = C ~d\alpha ~\sin\beta ~d\beta ~d\gamma,
\end{equation}
where $C$ is an arbitrary constant; since this will cancel out in the averages we set $C=1$.  In the probability the $\beta$ terms and the $\alpha+\gamma$ terms factorize, so that $\langle \sin\beta \cos\gamma \rangle = \langle \sin\beta \rangle \langle \cos\gamma \rangle$ and $\langle \cos\beta \sin\alpha \sin\gamma \rangle = \langle \cos\beta \rangle\langle \sin\alpha \sin\gamma \rangle$ etc. Also we find that $\langle \sin\alpha \rangle =\langle \sin\gamma \rangle=\langle \cos\alpha \rangle=\langle\cos\gamma \rangle$=0, and $\langle \sin\alpha\cos\gamma\rangle = \langle \cos\alpha\sin\gamma\rangle =0$. This leaves a diagonal matrix
\begin{widetext}
\begin{equation} \label{transfer_mean}
\langle \mathcal{R}\rangle =
\left( \begin{matrix} 
\langle\cos\beta\rangle & 0 & 0 \\
0  & \langle \cos\gamma\cos\alpha \rangle - \langle\cos\beta\rangle \langle \sin\alpha\sin\gamma \rangle& 0\\
 & 0 & - \langle \sin\alpha\sin\gamma \rangle + \langle\cos\beta\rangle \langle \cos\gamma\cos\alpha \rangle
\end{matrix} \right).
\end{equation}

The partition function for a segment is
\begin{align}
Z&= \int_{-\pi}^{\pi} d\alpha \int_0^{\pi} d\beta \int_{-\pi}^{\pi} d\gamma ~
 \sin\beta~\exp\left[-  \frac{\kappa_b}{a k_BT}(1-\cos\beta) -  \frac{\kappa_t}{a k_BT}\left[ 1-\cos(\alpha+\gamma) \right] \right]  \nonumber \\
&= \frac{2 k_BT a}{\kappa_b} e^{-\kappa_b/a k_BT} \sinh\left( \frac{\kappa_b}{a k_BT} \right) 
4\pi^2  e^{-\kappa_t/a k_BT} I_0\left( \frac{\kappa_t}{a k_BT} \right),
\end{align}
where $I_\nu(x)$ are the modified Bessel functions of the first kind.
\end{widetext}
Evaluating the remaining matrix elements we find
\begin{align}
\langle \cos\beta \rangle &= \frac{1}{\tanh\left( \kappa_b/a k_BT \right)} - \frac{a k_BT}{\kappa_b}, \nonumber \\
\langle \cos\gamma\cos\alpha \rangle &= \frac{I_1(\kappa_t/a k_BT)}{2 I_0(\kappa_t/a k_BT)},\nonumber \\
\langle \sin\alpha\sin\gamma \rangle &= - \frac{I_1(\kappa_t/a k_BT)}{2 I_0(\kappa_t/a k_BT)}.\nonumber
\end{align}
To get analytic expressions for these averages we expand about $a k_BT/\kappa_b\to0$ and $a k_BT/\kappa_t\to0$. To first order this gives
\begin{align}
\langle \cos\beta \rangle &=1-\frac{ak_BT}{\kappa_b} , \label{mncosbetalim}\\
\langle \cos\gamma\cos\alpha \rangle =-\langle \sin\alpha\sin\gamma \rangle&=\frac{1}{2}-\frac{a k_BT}{4\kappa_t}.\label{mncosalphacosgammalim}
\end{align}

The vectors $\unit{u}_i$, $\unit{f}_i$ and $\unit{v}_i$ are eigenvectors of $\langle \mathcal{R} \rangle$. The eigenvalue corresponding to $\unit{u}_i$ is $\langle \cos\beta \rangle$, meaning that 
\[
\langle \unit{u}_n \cdot \unit{u}_1 \rangle = \langle \cos\beta \rangle^{n-1} = e^{-a(n-1)/\xi_u},
\]
i.e. an exponential decay with a correlation length $\xi_u=-a/\ln \langle \cos\beta \rangle$. Using Eq.~(\ref{mncosbetalim}) gives 
\begin{equation}\label{xi_b}
\xi_u=\frac{\kappa_b}{k_BT}.
\end{equation}
We identify this as the bending persistence length (which is the same as in the continuous case -- see Appendix~\ref{app:corfun}).

The eigenvalue corresponding to the eigenvector $\unit{f}_i$ is $\langle \cos\alpha\cos\gamma \rangle (1+\langle \cos\beta \rangle)$. In a similar fashion to above, we can say that 
\begin{equation}
\langle \unit{f}_n \cdot \unit{f}_1 \rangle = [ \langle \cos\alpha\cos\gamma \rangle (1+\langle \cos\beta \rangle) ]^{n-1} = e^{-a(n-1)/\xi_f},
\end{equation}
where $\xi_f=-a/\ln[ \langle \cos\alpha\cos\gamma \rangle (1+\langle \cos\beta \rangle) ]$ is the $\unit{f}_i$ correlation length. Using Eqs.~(\ref{mncosbetalim}) and (\ref{mncosalphacosgammalim}) and taking the small $a$ limit gives 
\begin{equation}
\xi_f= \frac{2}{k_BT} \frac{\kappa_b \kappa_t}{\kappa_b+\kappa_t}.
\end{equation}
Note that in the limit of large $\kappa_b$, i.e. a straight rod, the twist correlation length reduces to $\xi_f=2\kappa_t/k_BT$; in other cases though, this correlation length is not such a useful quantity as it measures both bending and twisting.

Other useful properties of the chain are the mean of the cosine and sine of the twist angle between a pair of segments $\langle \cos(\alpha+\gamma) \rangle$ and $\langle \sin(\alpha+\gamma) \rangle$, and also the correlation in the total twist between two points along the chain. The first two quantities are found to be
\begin{align} \label{avCosab}
\langle \cos(\alpha+\gamma) \rangle&=\frac{I_1(\kappa_t/a k_BT)}{I_0(\kappa_t/a k_BT)}, \nonumber \\
& \approx 1-\frac{a k_BT}{2\kappa_t},
\end{align}
and
\begin{equation}\label{avSinab}
\langle \sin(\alpha+\gamma) \rangle=0
\end{equation}
The latter is the mean of the cosine of the total twist between two points, $\langle \cos\Omega_{n} \rangle$, where $\Omega_{n}= \sum_{i=1}^{n} (\alpha_i+\gamma_i)$ is the sum of twist angles between each of the beads from $1$ to $n$. Since $\langle \cos\Omega_{n} \rangle$ will decrease as $n$ increases, we can write $\langle \cos\Omega_{n} \rangle = e^{-na/\xi_{\rm Tw}}$, where $\xi_{\rm Tw}$ the twist correlation length. The average is given by
\begin{align}
\langle \cos \Omega_{n} \rangle &= 
\frac{\int \cos \left( \sum_{i=1}^n \Phi_i \right) \exp[-\sum_{i=1}^N E_i/k_BT] d\Gamma }
{\int  \exp[-\sum_{i=1}^N E_i/k_BT] d\Gamma} 
\nonumber \\
&= 
\frac{\int \exp[ \mathbbm{i} \sum_{i=1}^n \Phi_i ] \exp[-\sum_{i=1}^N E_i/k_BT] d\Gamma}
{2 \int  \exp[-\sum_{i=1}^N E_i/k_BT] d\Gamma} 
\nonumber \\&~+
\frac{\int \exp[ -\mathbbm{i} \sum_{i=1}^n \Phi_i ] \exp[-\sum_{i=1}^N E_i/k_BT] d\Gamma}
{2 \int  \exp[-\sum_{i=1}^N E_i/k_BT] d\Gamma} \nonumber
\end{align}
where $d\Gamma=\prod_{i=1}^N \mathcal{D}(\alpha_i,\beta_i,\gamma_i)$, and $\Phi_i=\alpha_i+\gamma_i$. Since the integral will be the same for each $i\leq n$, and terms for $i>n$ cancel, this simplifies to
\begin{align}
\langle \cos \Omega_n \rangle &= \frac{1}{2} \left[ 
\frac{\int e^{\mathbbm{i} \Phi} e^{-E/k_BT} \mathcal{D}(\alpha,\beta,\gamma)}
{\int  e^{-E/k_BT} \mathcal{D}(\alpha,\beta,\gamma)} 
\right]^n \nonumber \\&~+ \frac{1}{2}\left[ 
\frac{\int e^{-\mathbbm{i} \Phi} e^{-E/k_BT} \mathcal{D}(\alpha,\beta,\gamma)}
{\int  e^{-E/k_BT} \mathcal{D}(\alpha,\beta,\gamma)} 
\right]^n \nonumber \\
&= \frac{1}{2} \langle e^{\mathbbm{i}\Phi} \rangle^n
+
\frac{1}{2} \langle e^{-\mathbbm{i}\Phi} \rangle^n, \nonumber
\end{align}
and
\begin{align}
\langle e^{\pm \mathbbm{i}\Phi} \rangle &= \langle \cos(\alpha+\gamma) \rangle
\pm \mathbbm{i} \langle \sin(\alpha+\gamma) \rangle. \nonumber
\end{align}
Using Eqs.~(\ref{avCosab}) and (\ref{avSinab}) above gives
\begin{equation}
\langle \cos \Omega_n \rangle =  \left( \frac{I_1(\kappa_t/a k_BT)}{I_0(\kappa_t/a k_BT)}  \right)^n.
\end{equation}
and we find $\xi_{\rm Tw}=-a/\log(1-ak_BT/2\kappa_t)$. Finally, for small $ak_BT/\kappa_t$
\begin{equation}\label{xi_Tw}
\xi_{\rm Tw}=\frac{2 \kappa_t}{k_BT}.
\end{equation}

\section{Implementation of elastic rod polymers in \textsc{LAMMPS}}\label{lammpsimp}

The model described above cannot be easily incorporated into pre-existing scalable software such as the \textsc{lammps} molecular dynamics solver. In Ref.~\cite{Chirico1994} Chirico and Langowski describe a Brownian dynamics simulation scheme for a bead-and-spring polymer model of DNA, where at each time step the positions of each bead are incremented by $\delta\mathrm{r}_i$ and the $\unit{f}_i$ vectors rotated by $\delta\Phi_i$ about the tangents $\mathbf{r}_{i+1}-\mathbf{r}_i$, according to the energy given in Eq.~(\ref{chiricolangowski}) (plus terms for stretching and excluded volume). At each step they also make a correction to ensure that the $\unit{f}_i$ vectors remain perpendicular to the tangents (i.e. $\unit{u}_i$ are aligned along the backbone); this correction prevents a straightforward implementation of this model in \textsc{lammps} and other multi-purpose molecular dynamics software. Here we describe two alternative models in which we add an additional term to the energy in place of this correction.

In order to describe a bead-and-spring polymer with torsional rigidity the beads must have an orientation as well as a position. In \textsc{lammps} this can be achieved in two ways: either by representing each bead by a rigid body (a collection of point ``atoms'' which move and rotate as a unit), or by a spherical atom which has position and orientation. 

\subsection{Model 1 : Using dihedral interactions and ``patchy'' beads.}

\begin{figure}
\includegraphics{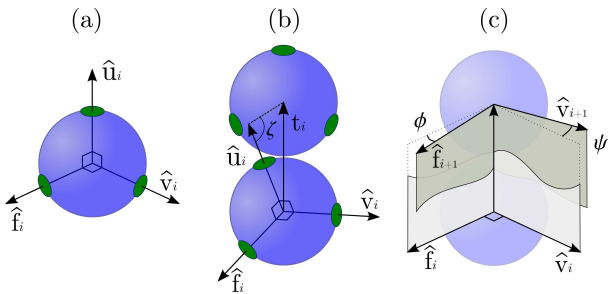}
\caption{\label{fig:dihedrals} Schematic diagram showing how three patches on a DNA bead can be used to describe the orientation of a DNA segment, according to Model 1 as detailed in the text. (a) The orientation of the $i$th bead is defined by a set of three ``patches'' rigidly fixed to the bead, which defines the set of unit axes vectors. (b) The tangent to the polymer at bead $i$ is defined $\mathbf{t}_i=\mathbf{r}_{i+1}-\mathbf{r}_i$, and the angle between $\unit{u}_i$ and $\mathbf{t}_i$ is denoted $\zeta$. (c) The twisting of the polymer is determined by dihedral angles $\phi$ and $\psi$ between planes as described in the text. }
\end{figure}

Consider a bead-and-spring polymer, where each bead is made up of a core sphere, and three small ``patches''. The position of the $i$th bead is denoted $\mathbf{r}_i$, and the positions of the patches on that bead are such that they lie along unit vectors $\unit{u}_i$, $\unit{f}_i$ and $\unit{v}_i$ which make up a right handed orthogonal set of axes [i.e. the bead and patches move as a unit such that the patch positions are $\mathbf{r}_i+(a/2)\unit{u}_i$ etc., where $a$ is the diameter of the bead --- see Fig.~\ref{fig:dihedrals}(a)].  Tangent vectors are defined $\mathbf{t}_i=\mathbf{r}_{i+1}-\mathbf{r}_i$. 

We write the energy of the system
\begin{align}\label{model1energy}
E_{\rm Model 1}=&\sum_{i=1}^{N-1} \Big[ \bar{\kappa}_b(1-\cos\theta_i) \nonumber \\
&+ \frac{\bar{\kappa}_t}{2}  ( 1 - \cos\phi_i )+ \frac{\bar{\kappa}_t}{2}  ( 1 - \cos\psi_i ) \nonumber \\
&+\bar{\kappa}_{a}(1-\cos \zeta_i) \Big] ,
\end{align}
where the first three terms are for bending and twisting, and the fourth is a term which keeps the orientation of the bead aligned with the backbone of the DNA; the energies $\bar{\kappa}_b$ and $\bar{\kappa}_t$ are the bending and twist rigidities scaled by the bond length ($\bar{\kappa}_b=\kappa_b/a$ etc.). The angle $\theta_i$ describes bending and is the angle between $\mathbf{t}_i$ and $\mathbf{t}_{i+1}$; $\phi_i$ is the angle between $\unit{f}_{i+1}$ and the plane defined by patch $f_i$, core $i$ and core $i+1$, and $\psi_i$ is the angle between $\unit{v}_{i+1}$ and the plane defined by patch $v_i$, core $i$ and core $i+1$. Finally, the angle $\zeta_i$ in the fourth term is the angle between core $i$, patch $u_i$ and core $i+1$. These angles can be written in terms of the unit vectors 
\begin{align}
\cos\theta_i &= \frac{ \mathbf{t}_i\cdot\mathbf{t}_{i+1} }{ |\mathbf{t}_i|~|\mathbf{t}_{i+1}| }, \label{costheta} \\
\cos\phi_i &= \frac{ (\mathbf{t}_i\times\unit{f}_i) \cdot (\mathbf{t}_i\times\unit{f}_{i+1})  }{ |\mathbf{t}_i\times\unit{f}_i| ~ |\mathbf{t}_i\times\unit{f}_{i+1}| }, \label{cosphi}\\
\cos\psi_i &= \frac{ (\mathbf{t}_i\times\unit{v}_i) \cdot (\mathbf{t}_i\times\unit{v}_{i+1})  }{ |\mathbf{t}_i\times\unit{v}_i| ~ |\mathbf{t}_i\times\unit{v}_{i+1}| },\label{cospsi} \\
\cos \zeta_i &= \unit{u}_i \cdot \frac{(\unit{u}_i-\mathbf{t}_i)}{|\mathbf{t}_i|}. \label{coszeta}
\end{align}
and are shown schematically in Figs~\ref{fig:dihedrals}(b) and (c).
This bead model and energy can be implemented in \textsc{lammps} by using the ``rigid body fix'' to integrate the dynamics of a patchy bead, angle interactions to implement the first and last terms in the energy, and dihedral interactions to implement the two twist terms (see \cite{lammps}).

In the limit of large $\kappa_{\rm a}$ the fourth term in Eq.~(\ref{model1energy}) will vanish since the angle $\zeta_i$ will approach zero and the unit vector $\unit{u}_i$ will be parallel with $\mathbf{t}_i$ --- i.e. $\unit{u}_i\to \mathbf{t}_i/|\mathbf{t}_i|$. The vectors $\unit{u}_i$, $\unit{f}_i$ and $\unit{v}_i$ then give a set of axis vectors which describe a frame of reference attached to the $i$th bead, and we recover the formalism used in Secs.~\ref{discrete} and {sec:corr}. Then Eq.~(\ref{costheta})-(\ref{cospsi}) can be written
\begin{align} \label{costhetaphipsi}
\cos\theta_i &=  \unit{u}_i\cdot\unit{u}_{i+1} ,\\
\cos\phi_i &= \frac{  (\unit{u}_i\times\unit{f}_{i+1}) \cdot \unit{v}_i  }{ |\unit{u}_i\times\unit{f}_{i+1}| },
\end{align}
and
\begin{align}\label{costhetaphipsi2}
\cos\psi_i = \frac{ -(\unit{u}_i\times\unit{v}_{i+1}) \cdot \unit{f}_i   }{ |\unit{u}_i\times\unit{v}_{i+1}| }.
\end{align}
As before the orientation of each bead can also be described by a set of Euler angles $\alpha_i$, $\beta_i$ and $\gamma_i$ which give the rotation which transforms the axes on bead $i$ into those on $i+1$, and we can use Eqs.~(\ref{vecip1i}) to relate these frames. Using this in Eqs.~(\ref{costhetaphipsi})-(\ref{costhetaphipsi2}) we get
\begin{align}
\theta_i &=  \beta_i ,\\
\cos\phi_i &= \frac{  \cos\gamma_i\cos\alpha_i- \cos\beta_i\sin\alpha_i\sin\gamma_i   }{ \sqrt{\cos^2\gamma_i + \cos^2\beta_i\sin^2\gamma_i} },\\
\cos\psi_i &= \frac{ \cos\beta_i\cos\alpha_i\cos\gamma_i-\sin\gamma_i\sin\alpha_i   }{ \sqrt{\sin^2\gamma_i + \cos^2\beta_i\cos^2\gamma_i}  }.
\end{align}
It is clear that the twist terms in Eq.~(\ref{model1energy}) do not exactly equal the term in the discrete elastic rod energy in Eq.~(\ref{Ediscreteeq}); to see if it is a good approximation we consider the limit of a stiff rod (small bending angles) and expand about $\beta_i\to 0$. To leading order in $\beta_i$ the cosines are
\begin{align}
\cos\phi_i &= \cos(\alpha_i+\gamma_i) \nonumber\\
&+ \frac{\beta_i^2}{2} \left( \sin\alpha_i \sin\gamma_i + \cos(\alpha_i+\gamma_i)\sin^2\gamma_i\right) +\mathcal{O}(\beta_i^4), \nonumber \\
\cos\psi_i &=\cos(\alpha_i+\gamma_i) \nonumber\\
&+ \frac{\beta_i^2}{2} \left( -\cos\alpha_i \cos\gamma_i + \cos(\alpha_i+\gamma_i)\cos^2\gamma_i \right)+\mathcal{O}(\beta_i^4). \nonumber
\end{align}
If we sum these the $\beta_i^2$ terms cancel, and we find that the twist term in Eq.~(\ref{model1energy}) is equal to that in Eq.~(\ref{Ediscreteeq}) up to the fourth order in $\beta_i$. An alternative model where only one dihedral is used (e.g. similar to those used in Refs.~\cite{Benedetti2013,Brackley2014}) would therefore not exactly recover Eq.~(\ref{Ediscreteeq}) to second order in $\beta_i$~\cite{foot_model1}.

In summary, this model can be implemented using the standard built in features of the \textsc{lammps} software, and up to fourth order in $\beta_i$ (and second order in $\zeta_i$~\cite{foot_zeta}) it will reproduce the discrete elastic rod model described by the Hamiltonian in Eq.~(\ref{Ediscreteeq}).

\subsection{Model 2 : Using spherical atoms.}\label{sec:model2}

An alternative to using patchy DNA beads and dihedral interactions is to use single spheres which have both a position and an orientation for the beads (achieved in \textsc{lammps} using the ``ellipsoid atom style''), and to enforce an orientational interaction between adjacent beads. Such an interaction is not available in the native \textsc{lammps} code, and we have implemented this as a new ``angle style''~\cite{sourcecode}. In a similar manner to previous sections we denote the position of each bead $\mathbf{r}_i$ and represent its orientation with three unit vectors $\unit{f}_i$, $\unit{v}_i$, and $\unit{u}_i$ which from a right-handed set of axes. Again tangent vectors are defined $\mathbf{t}_i=\mathbf{r}_{i+1}-\mathbf{r}_i$, and the Euler angles $\alpha_i$, $\beta_i$ and $\gamma_i$ give the orientation of bead $i+1$ with respect to bead $i$. 

We write down bending, twist and alignment terms in the energy
\begin{align}\label{Emodel2}
E_{\rm Model 2}=&  \sum_{i=1}^{N-1} \Big[ \bar{\kappa}_b(1-\cos\theta_i) 
+\bar{\kappa}_t  \left[ 1-\cos(\alpha_i+\gamma_i) \right] \nonumber \\
&+\bar{\kappa}_a  ( 1 - \cos\psi_i ) \Big],
\end{align}
where the third term acts to align the $\unit{u}_i$ vectors along the backbone of the polymer. It is straightforward to see that in the limit of large $\kappa_a$ the alignment term will vanish, $\unit{u}_i$ will be parallel to the tangent $\mathbf{t}_i$, and we recover the energy for the discrete elastic rod give in Eq.~(\ref{Ediscreteeq}).

In summary, implementation of Model 2 in the \textsc{lammps} code requires a new orientation angle style which is not part of the core software package. A derivation of the force and torque on each bead which results from this potential is given in Appendix~\ref{app:force}. The model reproduces the discrete elastic rod model [Eq.~(\ref{Ediscreteeq})] up to a correction which is second order in the angle $\psi_i$. 

\section{Comparing theory with numerical results.}

We now compare the theoretical results for the correlation functions derived in Sec.~\ref{sec:corr} with those from numerical simulations of bead-and-spring polymers. We use the \textsc{lammps} software to implement each of the models described in the previous section.

\begin{figure}
\includegraphics{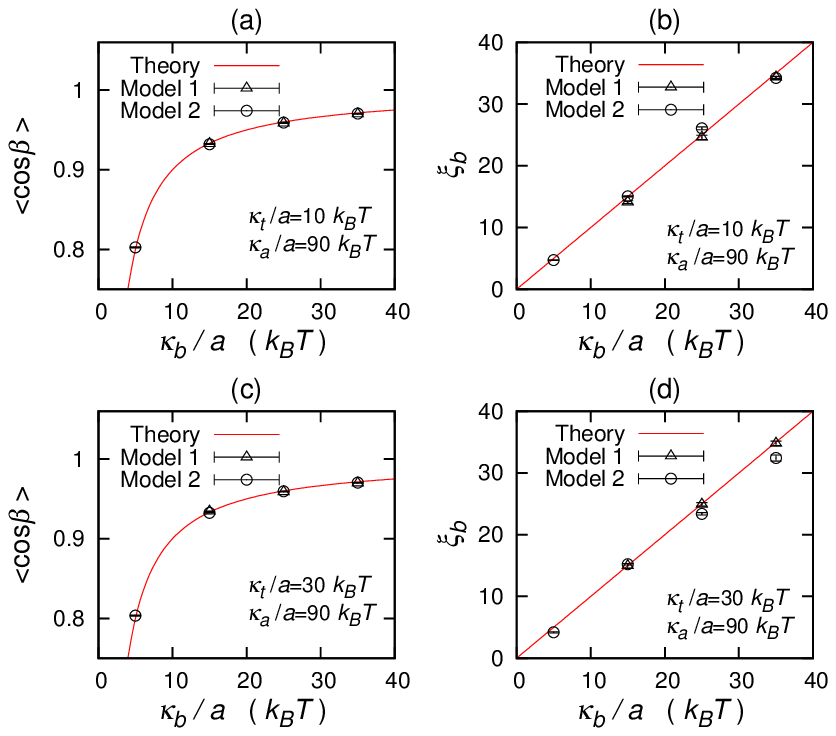}
\caption{\label{fig:data_bend} Plots comparing theory and simulations for the mean bending angle per segment and the bending correlation length. Parameters used are shown on the plots. (a) and (c) show $\langle\cos\beta\rangle$ with solid lines showing the theory as given in Eq.~(\ref{mncosbetalim}). (b) and (d) show the bending correlation length $\xi_u$ with solid lines showing the theory as given in Eq.~(\ref{xi_b}).}
\end{figure}

\begin{figure}
\includegraphics{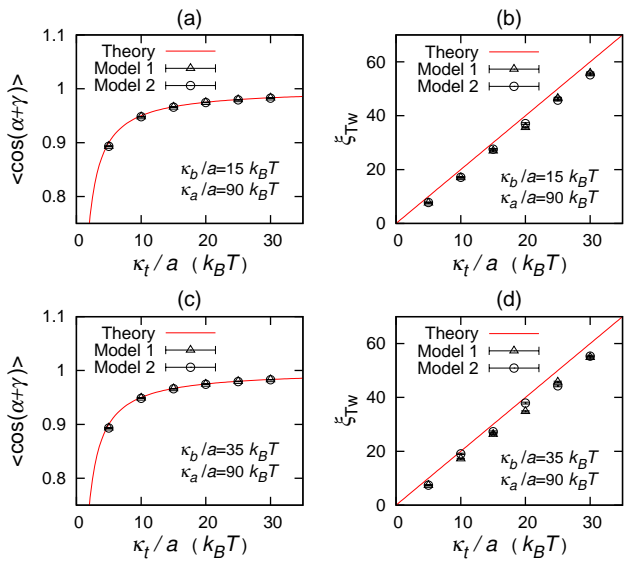}
\caption{\label{fig:data_twist}  Plots comparing theory and simulations for the mean twist per segment and the twist correlation length. Parameters used are shown on the plots. (a) and (c) show $\langle\cos(\alpha+\gamma)\rangle$ with solid lines showing the theory as given in Eq.~(\ref{avCosab}). (b) and (d) show the twist correlation length $\xi_{\rm Tw}$ with solid lines showing the theory as given in Eq.~(\ref{xi_Tw}).}
\end{figure}

Simulations were of a bead-and-spring polymer of length $N=200$ beads. The force fields used correspond either to the Hamiltonian given in  Eq.~(\ref{model1energy}) or to that in Eq.~(\ref{Emodel2}), plus FENE spring interactions between the beads, and shifted, truncated Lennard-Jones interactions, which were used to stop the beads from overlapping. Dynamics were evolved according to the Langevin equation (a scheme commonly know as Langevin dynamics, or Brownian dynamics with no hydrodynamic interactions). Averages and correlation lengths are measured from equilibrium configurations. 

In Fig.~\ref{fig:data_bend} the average bending angle and the bending persistence length are shown; for both models there is good agreement with theory. In Fig.~\ref{fig:data_twist} we show results for the twisting of the chain. This time there is a good agreement between theory and both models for averages over the individual twist angles ($\alpha+\gamma$). The long range correlation results agree less well, with the measured twist correlation length $\xi_{\rm Tw}$ slightly shorter than suggested by the theory. A likely source of this discrepancy is the fact that the Hamiltonian from which we obtain the expression for $\xi_{\rm Tw}$ does not contain the term in $\bar{\kappa}_a$ from the alignment interaction which is present in the simulations.

\section{Conclusions}

To summarise, we have discussed here a number of possible models for a 
twistable worm-like chain, which are suitable for use in coarse grained 
Brownian dynamics simulations. 
We have provided a detailed analysis of the connection between discretised 
and continuous energy functions, which shows that both standard formulations 
in terms of Euler angles (such as the one originally introduced in 
Ref.~\cite{Chirico1994}), and another description based on dihedrals, are 
equivalent with the continuous theory up to second order in the bending and 
twist angles. 
This link is not commonly discussed in the literature, yet it is crucial to 
prove that the theory has the right continuous limit. In our experience it
is quite easy to build up an energy function for twistable elastic rods which 
has the right behaviour for straight fibres, but does not obey the correct 
continuum theory when bending and twist fluctuations are simultaneously 
incorporated. Our explicit analysis provides a simple criterion to 
discriminate between viable and non-viable theories.
Furthermore, we have shown that, as expected, all ``correct'' models possess
well-behaved tangent-tangent and twist correlation functions, associated
with two different persistence lengths.

The continuum description of an elastic rod in Sec.~\ref{continuous} has been formulated for a linear rod with unconstrained ends. Several previous studies~\cite{Bouchiat2000,Moroz1997} have focused on supercoiled DNA by considering a rod stretched between two substrates; there the rod is parametrised by the deviation of its configuration from a reference relaxed molecule, an extra stretching term is included in the energy, and a constraint of constant linking number is imposed. That description suffers from a pathology in the continuum limit which arises because twisting a constrained rod will lead to writhing, and, unless self-avoidance is accounted for, it is possible for loops to form and the rod to pass through itself (this manifests as a singularity in the constant linking number Hamiltonian, and is related to the analogous system of a quantum mechanical symmetric top~\cite{Bouchiat2000}). 
This problem does not arise in our analytical treatment, because we do not need to employ any approximate formulas for writhe and linking number, as
we do not include supercoiling (since the ends of the rod are not constrained).
A similar pathology appears in the case where the energy includes a term in the geometric normal $\unit{n}$ (defined in Sec.~\ref{continuous}) instead of the material normal $\unit{f}$~\cite{transfermatrix}, which gives rise to complex eigenvalues of the transfer matrix; this results in the tangent-tangent correlations decaying in an oscillatory way, with the oscillation period becoming singular in the continuum limit. We do not encounter this problem in the present work, since the energy in Eq.~(\ref{Ediscreteeq}) leads to a transfer matrix [Eq.~(\ref{transfer_mean})] with real eigenvalues.

The main contribution of our work is that we have provided a detailed
implementation of the model of Refs.~\cite{Chirico1994} in LAMMPS~\cite{lammps},
a very well used code to simulate Brownian dynamics. The associated
code, which is available on request, allows the inclusion of Hamiltonians
defined in terms of Euler angles, and we expect it will be of use 
in the future to perform large scale and parallel coarse grained simulations of
DNA and of other polymers or fibres where twist, as well as twist fluctuations, 
play a major role in the physics (for some examples, see e.g. Refs.~\cite{powers,powers2}). 

In particular, one may use the implementation we have described to
study the dynamics and physics of confined supercoiled DNA, which is
a good starting point to describe bacterial DNA; it would also be 
possible, for instance, to model supercoiling in chromatin fibres and
chromosome fragments, to begin understanding its role in gene
regulation. The advantage of the approach we are proposing here is that,
thanks to the high scalability of LAMMPS, these simulations can reach
unprecedented large scale with respect to previous simulations
of coarse grained supercoiled DNA presented.

\begin{acknowledgments}
We acknowledge EPSRC grants EP/I034661/1 and EP/I004262/1 for funding.
\end{acknowledgments}

\appendix

\section{Derivation of correlation functions for the continuous elastic rod}\label{app:corfun}

Here we present an approximate but intuitive way of calculating the correlation functions associated with bending and twisting of an elastic rod. The same results can be obtained in a formal way by either performing corresponding path integrals~\cite{kamien} or solving a Fokker-Planck equation~\cite{fredrickson}.

Consider two close cross-sections positioned at $s$ and $s+\Delta s$ along the backbone of the rod, i.e. $\Delta s$ is small in some sense. A fluctuation resulting in a small rotation of the cross-section $s+\Delta s$ with respect to the $s$-cross-section is associated with an energy penalty which is given by Eq.~(\ref{e_er_omegas})
\begin{eqnarray}
&&\Delta E_{\rm ER} \nonumber \\
&&\quad = \frac{1}{2} \int_s^{s+\Delta s} \left[ \kappa_{b}\left(\omega_1(s)^2 + \omega_2(s)^2 \right)+ \kappa_t\omega_3(s)^2 \right] ds \nonumber \\
&&\quad \approx \frac{\Delta s}{2} \left[ \kappa_{b}\left(\omega_1^2 + \omega_2^2 \right)+ \kappa_t\omega_3^2 \right],
\end{eqnarray}
where $\omega_1$, $\omega_2$ and $\omega_3$ are the rotation \emph{rates}, i.e. changes in the angles per $\Delta s$, defining the orientation of the $s+\Delta s$-cross-section with respect to the $s$-cross-section. The probability distribution for the $\omega$'s is then given by
\begin{eqnarray}
&&P\left(\omega_1,\omega_2,\omega_3\right) \label{probdist} \\
&&\quad = P_0 \exp{\Big\{-\frac{\Delta s}{2 k_B T} \left[ \kappa_{b}\left(\omega_1^2 + \omega_2^2 \right)+ \kappa_t\omega_3^2 \right]\Big\}},\nonumber
\end{eqnarray}
where $k_B$ is the Boltzmann constant, $T$ is the temperature, and the normalisation constant $P_0$ is obtained by integrating Eq.~(\ref{probdist}) over the permitted values of $\omega$'s. We observe that the integrand in this and all similar integrals is sharply peaked around $\omega_1=\omega_2=\omega_3=0$, and, therefore, we can replace the true limits of integration with an infinite range, thus greatly simplifying further analysis. 

The correlation function $\langle \unit{u}(s) \cdot \unit{u}(s+\Delta s) \rangle$, describing propagation of bending along the rod, can now be evaluated
\begin{eqnarray}
&\langle \unit{u}(s) \cdot \unit{u}(s+\Delta s) \rangle \nonumber \\
&\approx \left\langle \unit{u}(s) \cdot \left( \unit{u}(s) + \unit{u}'(s)\Delta s + \unit{u}''(s)\frac{\Delta s^2}{2} + \cdots\right) \right\rangle \nonumber \\
&= 1-\left\langle \frac{\Delta s^2}{2}\left(\omega_1^2 + \omega_2^2\right) \right\rangle,
\label{uu}
\end{eqnarray}
where the average is taken with respect to the probability distribution Eq.~(\ref{probdist}) and we have used Eq.~(\ref{GFSeq}) to calculate $\unit{u}'(s)$ and $\unit{u}''(s)$; here primes denote derivatives with respect to $s$. Evaluating the Gaussian integral in Eq.(\ref{uu}), we obtain
\begin{eqnarray}
1-\left\langle \frac{\Delta s^2}{2}\left(\omega_1^2 + \omega_2^2\right) \right\rangle = 1 - \frac{\Delta s}{\lambda_u}\approx e^{-\Delta s/\lambda_u},
\end{eqnarray}
where the last equality holds in view of smallness of $\Delta s$, and $\lambda_u = \kappa_{b}/k_B T$ is the bending correlation length of the rod.

In a similar fashion, it can be shown that
\begin{eqnarray}
&\langle \unit{f}(s) \cdot \unit{f}(s+\Delta s) \rangle \nonumber \\
&\approx 1-\left\langle \frac{\Delta s^2}{2}\left(\omega_2^2 + \omega_3^2\right) \right\rangle \approx e^{-\Delta s/\lambda_f},
\label{ff}
\end{eqnarray}
where $\lambda_f = [2/k_BT][\kappa_b\kappa_t/(\kappa_b+\kappa_t)]$.

\section{Implementation of Orientation Interaction in \textsc{lammps}}\label{app:force}

To implement Model 2 [described in Sec.~\ref{sec:model2}] in \textsc{lammps} we have written a new ``angle style'', which adds an orientation interaction between two adjacent DNA beads. This incorporates forces and torques which originate from the twist and alignment terms in Eq.~(\ref{Emodel2}), which we denote
\begin{align}
U= \kappa_t \sum_i &\left( 1-\cos(\alpha_i+\gamma_i) \right) \nonumber \\
&+\kappa_a \sum_i ( 1 - \cos\psi_i ), \label{Utors}
\end{align}

Below we first derive the force and torque which result from these terms, and then describe how this new angle style can be used in \textsc{lammps}.

\subsection*{Force and Torque}

We proceed in a similar manner to the derivation in Ref.~\cite{Chirico1994}. Some useful expressions linking the Euler angles and the orientation vectors are
\begin{equation}
\unit{u}_{i+1}\cdot\unit{u}_i = \cos\beta_i,
\end{equation}
and
\begin{equation}\label{cosalpha+gamma}
\unit{f}_{i+1}\cdot\unit{f}_i + \unit{v}_{i+1}\cdot\unit{v}_i = (1+\unit{u}_{i+1}\cdot\unit{u}_i) \cos\left(\alpha_i+\gamma_i\right).
\end{equation}
The angle between the vectors $\mathbf{t}_i$ and $\mathbf{t}_{i+1}$ is denoted $\theta_i$, and the angle between vectors $\unit{u}_i$ and $\mathbf{t}_i$ is denoted $\psi_i$, and these are given by
\begin{equation}\label{costhetaidentity}
\cos\theta_i=\frac{ \mathbf{t}_i\cdot\mathbf{t}_{i+1} }{ b_i b_{i+1} }
\end{equation}
and
\begin{equation}\label{cosphiidentity}
b_i \cos\psi_i=\unit{u}_i\cdot\mathbf{t}_i,
\end{equation}
where $b_i=|\mathbf{t}_i|$ (note that in Sec.~\ref{discrete} we considered a discrete rod with fixed bond length $b_i=a$, but in simulations a spring potential is used to constrain the bond length, so the $b_i$ will be distributed about a mean value). Consider that in a time $\de{t}$ the position and orientation of the $i$th bead change by $\de{\mathbf{r}}_i$ and $\de{\phi}_i$ respectively, where the latter is a rotation about an axis $\unit{p}_i$. Assuming that $\left\{ \de{\mathbf{r}}_i,\de{\phi}_i \right\}$ are independent variables, then via the principal of virtual work the force and torque on each bead due to the potential $U$ is given by
\begin{equation}\label{vwork}
\de{U} = - \sum_i \mathbf{F}_i\cdot\de{\mathbf{r}}_i - \sum_i \mathbf{T}_i\cdot\unit{p}_i \de{\phi}_i.
\end{equation}
To find $\mathbf{F}_i$ and $\mathbf{T}_i$ we can take the derivative of Eq. (\ref{Utors})
\begin{align}\label{diffUtors}
\de{U} = \kappa_t \sum_i &\sin(\alpha_i+\gamma_i) \de{(\alpha_i+\gamma_i)} \nonumber \\
&+ \kappa_a \sum_i \sin\psi_i \de{\psi}_i,
\end{align}
and then equate terms in $\de{\mathbf{r}}_i$ and $\de{\phi}_i$ in Eq.~(\ref{vwork}). 

Differentiating Eq. (\ref{cosphiidentity}) gives
\begin{align}\label{diffcosphi}
\cos\psi_i \de{b}_i - b_i \sin\psi_i\de{\psi}_i &= \nonumber \\ \de{\unit{u}}_i\cdot\mathbf{t}_i &+ \unit{u}_i \cdot \left( \de{\mathbf{r}_{i+1}} - \de{\mathbf{r}}_i \right),
\end{align}
and $\de{b}_i$ can be found by differentiating $b_i^2=\mathbf{t}_i\cdot\mathbf{t}_i$ to give 
\[
b_i\de{b}_i=\mathbf{t}_i\cdot \left( \de{\mathbf{r}}_{i+1} - \de{\mathbf{r}}_i \right).
\]
The infinitesimal change in the axis vector $\unit{u}_i$ due to a rotation of $\de{\phi}_i$ about a vector $\unit{p}_i$ is given by $\de{\unit{u}}_i=\unit{p}_i\times\unit{u}_i\de{\phi}_i$~\cite{foot_rotmat}. Equation~(\ref{diffcosphi}) can then be written
\begin{align}\label{finaldiffcosphi}
\sin\psi_i\de{\psi}_i = 
\frac{1}{b_i} &\left( \cos\psi_i \unit{t}_i - \unit{u}_i \right) \cdot \left( \de{\mathbf{r}}_{i+1} - \de{\mathbf{r}}_i \right) \nonumber \\
&- (\unit{u}_i\times\unit{t}_i) \cdot \unit{p}_i \de{\phi}_i,
\end{align}
where $\unit{t}_i=\mathbf{t}_i/b_i$.

Differentiating Eq. (\ref{cosalpha+gamma}) gives
\begin{widetext}
\begin{align}
\cos\left(\alpha_i+\gamma_i\right) \left[ \de{\unit{u}}_{i+1}\cdot\unit{u}_i + \unit{u}_{i+1}\cdot\de{\unit{u}}_i \right] - \sin\left(\alpha_i+\gamma_i\right) \left[ 1+\unit{u}_{i+1}\cdot\unit{u}_i \right] \de{\left(\alpha_i+\gamma_1\right)} = \nonumber \\ 
\de{\unit{f}}_{i+1}\cdot\unit{f}_i + \unit{f}_{i+1}\cdot\de{\unit{f}}_i  +
\de{\unit{v}}_{i+1}\cdot\unit{v}_i + \unit{v}_{i+1}\cdot\de{\unit{v}}_i, \nonumber
\end{align}
which, again using $\de{\unit{u}}_i=\unit{p}_i\times\unit{u}_i\de{\phi}_i$ and similar for $\unit{f}_i$, $\unit{v}_i$ etc., can be re-written
\begin{align}\label{finaldiffcosalpha+gamma}
\left[ 1+\unit{u}_{i+1}\cdot\unit{u}_i \right] & \sin\left(\alpha_i+\gamma_1\right) \de{\left(\alpha_i+\gamma_1\right)} = \nonumber \\ 
&\left( (\unit{u}_{i+1}\times\unit{u}_i) \cos\left(\alpha_i+\gamma_1\right) - (\unit{f}_{i+1}\times\unit{f}_i) - (\unit{v}_{i+1}\times\unit{v}_i) \right) \cdot
\left( \unit{p}_{i+1}\de{\phi}_{i+1} - \unit{p}_i \de{\phi}_i \right).
\end{align}
\end{widetext}

Inserting Eqs.~(\ref{finaldiffcosphi}) and (\ref{finaldiffcosalpha+gamma}) into Eq.~(\ref{diffUtors}), and matching terms in Eq.~(\ref{vwork}) gives expressions for the force and torque
\begin{align}
\mathbf{F}_i &= \kappa_a \left( \mathbf{G}_i - \mathbf{G}_{i-1} \right),
\nonumber \\
\mathbf{T}_i &= \kappa_a \unit{u}_i\times\unit{t}_i + \kappa_t \left( \mathbf{H}_i - \mathbf{H}_{i-1} \right),
\end{align}
where
\begin{equation}
\mathbf{G}_i = \frac{1}{b_i} \left[ (\unit{u}_i\cdot\unit{t}_i) \unit{t}_i - \unit{u}_i \right],
\end{equation}
and
\begin{equation}
\mathbf{H}_i = \frac{ \unit{u}_{i+1}\times\unit{u}_i \cos\left(\alpha_i+\gamma_1\right) - \unit{f}_{i+1}\times\unit{f}_i - \unit{v}_{i+1}\times\unit{v}_i }{ 1+\unit{u}_{i+1}\cdot\unit{u}_i }.
\end{equation}

For a circular polymer with $N$ beads the sums over $i$ in Eqs.~(\ref{Utors}) and (\ref{vwork}) are from $1\to N$, and $\alpha_N$, $\beta_N$ and $\gamma_N$ are the Euler angles describing the rotation to orientate bead $N$ to bead 1,  $\mathbf{t}_N=\mathbf{r}_1-\mathbf{r}_N$, and $\theta_N$ is the angle between $\mathbf{t}_N$ and $\mathbf{t}_1$. For a linear polymer $\alpha_N$, $\beta_N$ $\gamma_N$, $\mathbf{t}_N$ and $\theta_N$ are not defined, so the sums run from $1\to N-1$; $\mathbf{F}_N$ and $\mathbf{T}_N$ therefore only have the $\mathbf{G}_{N-1}$ and $\mathbf{H}_{N-1}$ terms in this case.

\subsection*{Use in LAMMPS}

We have implemented two new \textsc{lammps} angle styles, \textit{polytors} and \textit{polytorsend}~\cite{sourcecode} to add the forces and torques derived above to a simulation of a linear or circular polymer. The procedure to set up the force field is as follows: (1) Use the atom style ``ellipsoids'' (atoms with orientation) and set up suitable initial conditions, e.g. beads in a random walk configuration. (2) Add bond interactions between adjacent pairs of beads, e.g. using harmonic or FENE bond styles. (3) Add angle interactions between adjacent triplets of beads to provide bending stiffness, e.g. using the cosine angle potential. (4) Add angle interactions of style \textit{polytors} between adjacent pairs of beads for the twist and alignment interactions. In \textsc{lammps} angle interactions specify three atoms; in the case of the \textit{polytors} style only two atoms are required, so the third atom id specified is ignored. For example a \textit{polytors} angle between beads 1 and 2 adds forces and torques which act (a) to align the $\unit{u}_1$ axis so that it points toward bead 2, and (b) to minimise the angle $\alpha_1+\gamma_1$. For a circular polymer include a \textit{polytors} angle interaction between beads $N$ and $1$; for a linear DNA include both a \textit{polytors} and a \textit{polytorsend} angle interaction between beads $N-1$ and $N$.


\begin{thebibliography}{99}

\bibitem{wlc1} J.~F. Marko and E.~D. Siggia, 
{\it Macromolecules} {\bf 28}, 8759 (1995).
\bibitem{wlc2} C. Bustamante, J.~F. Marko, E.~D. Siggia and S. Smith,
{\it Science} {\bf 265}, 1599 (1994).
\bibitem{wlc3} P.~L. Hansen and R. Podgornik, 
{\it J. Chem. Phys.} {\bf 114}, 8637 (2001).
\bibitem{wlc4} A. Rosa, T.~X. Hoang, D. Marenduzzo and A. Maritan,
{\it Macromolecules} {\bf 36}, 10095 (2003).
\bibitem{wlc5} N.~M. Toan, D. Marenduzzo, C. Micheletti, 
{\it Biophys. J.} {\bf 89}, 80 (2005).
\bibitem{understandingdna} C.~R. Calladine and H. Drew, 
{\it Understanding DNA}, Academic Press, San Diego (1992).
\bibitem{marko1} J.~F. Marko and E.~D. Siggia, {\it Science}
{\bf 265}, 506 (1994).
\bibitem{marko3} J.~F. Marko and E.~D. Siggia,
{\it Macromolecules} {\bf 27}, 981 (1994).
\bibitem{supercoilingreview} N. Gilbert and J. Allan,
{\it Curr. Opin. Gen. Devel.} {\bf 25}, 15 (2014).
\bibitem{marko2} J.~F. Marko and E.~D. Siggia, {\it Phys. Rev. E}
{\bf 52}, 2912 (1995).
\bibitem{Forth2008} S. Forth, C. Deufel, M.Y. Sheinin, B. Daniels, J.P. Sethna, and M.D. Wang, {\it Phys. Rev. Lett.} {\bf 100} 148301 (2008).
\bibitem{bacterialsupercoiling1}
L. Postow, C.~D. Hardy, J. Arsuaga and N.~R. Cozzarelli, 
{\it Gen. Devel.} {\bf 18}, 1766 (2004).
\bibitem{mirnybacteria} T.~B.~K. Le, M.~V. Imakaev, L.~A. Mirny and
M.~T. Laub, {\it Science} {\bf 342}, 731 (2013).
\bibitem{twinsupercoiled} L.~F. Liu and J.~C. Wang, {\it Proc. Natl. Acad.
Sci. USA} {\bf 84}, 7024 (1987).
\bibitem{transcription} Y.~P. Tsao, H.~Y. Wu and L.~F. Liu,
{\it Cell.} {\bf 56}, 111 (1989).
\bibitem{Deng2005} S. Deng, R.A.  Stein, and N.P. Higgins, {\it Molecular Microbiology} {\bf 57} 1511 (2005).
\bibitem{twinsupercoiledBD} S.~P. Mielke, W.~H. Fink, V.~V. Krishnan, 
N. Groenbech-Jensen and C.~J. Benham, {\it J. Chem. Phys.}
{\bf 121}, 8104 (2004).
\bibitem{supercoilingnick} C. Naughton {\it et al.},
{\it Nat. Struct. Mol. Biol.} {\bf 20}, 387 (2013).
\bibitem{supercoilingchromatin} D.~J. Clark and G. Felsenfeld,
{\it EMBO J.} {\bf 10}, 387 (1991).
\bibitem{Roca2011} J. Roca, {\it Chromosoma} {\bf 120} 323 (2011).
\bibitem{Mondal2001} N. Mondal and J.D. Parvin, {\it Nature} {\bf 413} 435 (2001).
\bibitem{Bouchiat2000} C. Bouchiat and M. M\'{e}zard, {\it Eur. Phys. J. E} {\bf 2} 377 (2000).
\bibitem{Fujimoto1990} B. S. Fujimoto and J. M. Schurr, {\it Nature} {\bf 344} 175 (1990).
\bibitem{Lipfert2010} J. Lipfert, J.W.J. Kerssemakers, T. Jager, and N.H. Dekker, {\it Nat Meth} {\bf 7} 977  (2010).
\bibitem{powersreview} T.~R. Powers, {\it Rev. Mod. Phys.} {\bf 82}, 1607 (2010).
\bibitem{twistsim1} W.~K. Olson, {\it Curr. Opin. 
Struct. Biol.} {\bf 6}, 242 (1996).
\bibitem{twistsim2} R.~K.~Z. Tan and S.~C. Harvey,
{\it J. Mol. Biol.} {\bf 205}, 573 (1989). 
\bibitem{twistsim3}  G.~C. Rollins, A.~S. Petrov, and S.~C. Harvey, 
{\it Biophys. J.} {\bf 94}, L38 (2008).
\bibitem{depablo} T.~A. Knotts, N. Rathore, D.~C. Schwartz and
J.~J. de Pablo, {\it J. Chem. Phys.} {\bf 126}, 084901 (2007).
\bibitem{schlick1} J. Huang, T. Schlick and A. Vologodskii, 
{\it Proc. Natl. Acad. Sci. USA} {\bf 98}, 968 (2001).
\bibitem{schlick2} T. Schlick and W.~K. Olson, {\it Science} {\bf 257}, 1110 (1992).
\bibitem{schlick3} T. Schlick, {\it Curr. Opin. Struc. Biol.} {\bf 5}, 245 (1995).
\bibitem{rappaport2} S. Rappaport and Y. Rabin, {\it Macromolecules} {\bf 37}, 7847 (2004).

\bibitem{Chirico1994} G. Chirico and J. Langowski {\it Biopolymers} 
{\bf 34}, 415 (1994).
\bibitem{Chirico1996} G. Chirico and J. Langowski {\it Biophys J.} {\bf 71}, 955 (1996).
\bibitem{Allison1989} S. Allison, R. Austin and Mike. Hogan,
{\it J. Chem. Phys.} {\bf 90}, 3843 (1989).
\bibitem{lammps} S.~J. Plimpton, {\it J. Comp. Phys.} {\bf 117}, 1 (1995) (http://lammps.sandia.gov).
\bibitem{Rappaport2007} S.~M. Rappaport and Y. Rabin {\it J. Phys. A} 
{\bf 40}, 4455 (2007).

\bibitem{foot_N} In the case of a linear rod there are $N$ position vectors describing the vertices of $N-1$ segments, with tangents $\unit{u}_i$ with $i=1,\dots,N-1$. For a circular rod there are $N$ segments and $\unit{u}_N=(\mathbf{r}_1-\mathbf{r}_N)/a$

\bibitem{transfermatrix} D. Marenduzzo, C. Micheletti, H. Seyed-allaei,
A. Trovato and A. Maritan, {\it J. Phys. A} {\bf 38}, L277 (2005).
\bibitem{Benedetti2013} F. Benedetti, J. Dorier, Y. Burnier,  and A. Stasiak, {\it Nucl. Aci. Res.} {\bf  42} 2848  (2013).
\bibitem{Brackley2014} C.A. Brackley, J. Allan, D. Keszenman-Pereyra and D. Marenduzzo, {\it Submitted} (2014).

\bibitem{foot_model1} The two potentials are however equivalent in the limit where all of $\alpha_i$, $\beta_i$ and $\gamma_i$ are small, a regime which is forced by the Hamiltonian considering either of the dihedrals.

\bibitem{foot_zeta} The angles $\zeta_i$ are minimised by choosing a large alignment energy $\kappa_a$; in practice this can be set as large as possible subject to numerical stability.

\bibitem{sourcecode} Source code for the \textsc{lammps} orientation angle style discussed in 
section~\ref{sec:model2} is available on request from the authors.

\bibitem{Moroz1997} J.~D. Moroz  and P. Nelson, 
{\it Proc. Nat. Acad. Sci.} {\bf 94}, 14418  (1997).

\bibitem{powers} S.~A. Koehler, T.~R. Powers, {\it Phys. Rev. Lett.}
{\bf 85}, 4827 (2000).
\bibitem{powers2} C.~W. Wolgemuth, T.~R. Powers and R.~E. Goldstein, 
{\it Phys. Rev. Lett.} {\bf 84}, 1623 (2000).

\bibitem{kamien} R.~D. Kamien, {\it Rev. Mod. Phys.} 
{\bf 74}, 953 (2002).
\bibitem{fredrickson} G.~H. Fredrickson, {\it The equilibrium theory
of inhomogeneous polymers}, Oxford University Press, New York (2006).

\bibitem{foot_rotmat} If $R$ is a rotation matrix for rotating a vector $\unit{u}$ by an angle $\de\phi$ about some axis $\unit{p}$, then $\de \unit{u}=(R-I)\unit{u}$. In the limit $\de\phi\to0$, $\unit{p}\times\unit{u}\de\phi=(R-I)\unit{u}$. 

\end{thebibliography}
\end{document}